\documentclass{article}
\usepackage[utf8]{inputenc}
\usepackage{kotex}

\title{인공지능 기반 공격 그래프 생성}
\author{고려대학교 정보보호대학원 해킹대응기술연구실 \\ (박상범, 이재성, 유정도, 송민근, 이효선, 최재웅, 사공채연, 김휘강)}
\date{November 2023}

\usepackage[numbers]{natbib}
\usepackage{graphicx}
\usepackage{subcaption}
\usepackage{amsmath}
\usepackage{amssymb}
\usepackage{verbatim}
\usepackage{cleveref}
\usepackage{titlesec}
\usepackage{makecell}
\usepackage{multirow}
\usepackage{multicol}
\usepackage{booktabs}
\usepackage{adjustbox}
\usepackage{times}
\usepackage{fancyhdr,graphicx,amsmath,amssymb}
\usepackage[ruled,vlined]{algorithm2e}

\titleformat{\paragraph}
{\normalfont\normalsize\bfseries}{\theparagraph}{1em}{}
\titlespacing*{\paragraph}
{0pt}{3.25ex plus 1ex minus .2ex}{1.5ex plus .2ex}

\begin{document}

\maketitle

\begin{abstract}
IoT 기술의 발전으로 많은 전자 기기들이 네트워크로 연결되어 서로 통신하고 특정 역할을 수행하고 있다. 그러나 네트워크에 많은 기기들이 연결되는 만큼 이로 인한 사이버 공격 위협도 늘어나고 있다. 사이버 공격 위협을 예방하고 탐지하는 것은 중요하며, 공격 위협을 예방하는 방법 중 하나로 공격 그래프를 사용할 수 있다. 공격 그래프는 네트워크의 보안 위협을 평가하는데 널리 사용되는 방법이다. 그러나 네트워크의 규모가 커질수록 공격 그래프를 생성하는데 걸리는 시간이 길어진다는 단점이 있다. 이 단점을 극복하기 위해 인공지능 모델을 이용할 수 있다. 인공지능 모델을 이용함으로써 짧은 시간 내에 공격 그래프를 만들며, 그 생성된 공격 그래프는 최적의 결과에 근사된 그래프이다. 공격 그래프 생성을 위한 인공지능 모델은 encoder와 decoder로 구성되며, 강화학습 알고리즘으로 학습된다. 인공지능 모델을 학습한 뒤 모델이 잘 학습되었는지 loss값과 reward 값의 변화를 확인하였다. 또한, 인공지능 모델이 생성한 공격 그래프가 기존의 방식으로 생성한 공격 그래프와 비슷한지 비교하였다.
\end{abstract}


\subsubsection*{Acknowledgment}
This work was supported by Institute of Information \& Communications Technology Planning \& Evaluation (IITP) grant funded by the Korea government (MSIT) (No. 2021-0-00624, Development of Intelligence Cyber Attack and Defense Analysis Framework for Increasing Security Level of C-ITS)

\section{서론}
오늘날 많은 PC, 노트북, 스마트폰 등 많은 전자 기기들이 네트워크로 연결되어 서로 통신하고 특정 역할을 수행하고 있다.
특히, IoT 기술의 발전은 수많은 기기들을 네트워크에 통합시키고 있는데, 이로 인한 사이버 공격 위협도 늘어나고 있다.
네트워크에 많은 기기들이 통신하고 있는 만큼 하나의 기기가 공격 받는 경우 네트워크를 통해 해당 기기에 연결된 다른 기기가 연달아 공격받을 수 있다.
따라서, 사이버 공격 위협을 예방하고 탐지하는 것은 연속적인 공격을 막는데 있어 중요하다.
사이버 공격 위협을 예방하는 방법 중 하나로 공격 그래프가 있다.

공격 그래프는 네트워크의 보안 위협을 평가하는데 널리 사용되는 방법이다.
공격 그래프에는 여러 유형이 있으며, 그 중 Bayesian attack graph (BAG)이 많이 사용되고 있다.
BAG에서 일반적으로 그래프의 노드는 PC, 노트북, IoT 기기, 스마트폰 등 네트워크에 연결된 시스템을 나타내며, 시스템 간의 연결 관계는 엣지를 나타낸다.
이때 취약점을 익스플로잇하여 한 노드에서 다른 노드로의 사이버 공격이 성공할 확률을 엣지의 가중치로 나타낸다.
공격 그래프는 한 시스템에서 다른 특정 시스템을 공격할 때 얼마나 공격이 feasible한지, 중간에 어떤 시스템을 거치는 것이 상대적으로 더 feasible한지 평가할 수 있다.
그러나 공격 그래프는 네트워크의 규모가 커지면 생성하는데 시간이 오래 걸린다는 단점이 있다.
네트워크에 포함된 시스템의 갯수가 늘어날 수록 (노드가 많을 수록), 시스템 간의 연결 갯수가 늘어날 수록 (엣지가 많을 수록) 공격 그래프를 생성하는데 걸리는 시간이 기하급수적으로 늘어나기 때문에 규모가 크거나 복잡한 네트워크에는 공격 그래프를 활용하기 어렵다.

이러한 단점을 해결하기 위해 인공지능 기반의 공격 그래프 생성을 제안한다.
기존의 공격 그래프의 생성 방식은 모든 노드와 엣지를 탐색하여 가능한 모든 공격 경로(Attack Path) 를 평가하므로 최적의 평가 결과를 정확하게 낼 수 있으나 생성 시간이 매우 길다.
제안하는 방법은 학습된 인공지능 모델을 이용해 공격 그래프를 생성하는 것으로, 인공지능 모델의 결과는 모델이 예측한 것이므로 기존의 생성 방식과는 달리 정확하지는 않지만 근사된 결과를 얻을 수 있으며, 생성 시간도 짧다.

본 기술문서는 다음과 같이 구성되어 있다.
Section \ref{sec:background}에서는 본 기술문서에서 기술하는 것과 관련된 개념과 용어를 설명하며, Section \ref{sec:literature}에서는 참고한 관련 연구를 소개한다.
Section \ref{sec:methodology}에서는 공격 그래프를 생성하기 위한 인공지능 모델을 학습하는 방법을 기술하고, Section \ref{sec:experiment}에서 해당 인공지능 모델의 실험 및 결과를 보인다.
Section \ref{sec:conclusion}에서는 본 기술문서의 내용을 정리하고 마무리한다.

\section{배경 및 용어}
\label{sec:background}
\subsection{공격 그래프}

``공격 그래프(Attack Graph)"는 컴퓨터 시스템에서 발생할 수 있는 다양한 보안 위협과 공격 경로(Attack Path)를 시각적으로 나타낸 그래픽 모델이다. 이 그래프는 보안 전문가들이 시스템의 취약점을 이용하여 시스템을 침투하고 공격할 수 있는 방법을 분석하고 이해하는 데 사용된다. 

공격 그래프는 일반적으로 네트워크 환경을 기반으로 하며, 시스템 내의 각 노드는 네트워크에 존재하는 시스템 또는 서비스를 나타낸다. 각 노드 간의 연결은 공격자가 시스템을 침투하기 위해 이용할 수 있는 특정 경로를 나타낸다. 이러한 시스템의 취약점, 서비스 구성 및 보안 정책 등을 기반으로 생성된다.

공격 그래프를 사용하면 보안 전문가들은 시스템 내의 취약점을 식별하고 공격 경로를 시각적으로 파악할 수 있다. 이를 통해 시스템 보안을 강화하고 취약점을 해결함으로써 공격 경로를 차단할 수 있다. 또한, 공격 그래프를 사용하면 보안 정책을 개선하거나 변경함으로써 시스템 내의 보안을 높일 수 있는 방법을 찾을 수 있다.
요약하자면, 공격 그래프는 시스템의 보안 취약점과 공격 경로를 시각적으로 표현한 도구로, 시스템 보안을 강화하고 보안 정책을 최적화하기 위해 사용된다.
\subsection{공격 경로}

공격 경로는 공격자가 시스템에 침투하고 공격을 수행하는 데 사용할 수 있는 특정한 경로나 순서를 나타냅니다. 이 경로는 여러 단계로 나뉘어져 있을 수 있고, 공격자는 각 단계에서 시스템의 취약점을 이용하여 시스템에 침투하거나 공격을 수행할 수 있다. 공격 경로를 이용하여 공격자는 시스템 내의 중요한 데이터나 리소스에 접근하거나 시스템을 제어할 수 있다. 
\subsection{GNN}

GNN은 Graph Neural Network의 약자로, 그래프 구조 데이터를 다루기 위한 신경망 모델을 나타낸다. 그래프는 노드와 엣지로 이루어진 네트워크 구조를 의미하며, 이러한 그래프 데이터는 다양한 분야에서 발생한다. 예를 들어, 소셜 네트워크, 분자 구조, 지식 그래프, 웹 페이지 간의 링크 구조 등이 그래프 형태의 데이터이다.

기존의 신경망 모델은 주로 이미지나 시퀀스 데이터와 같은 구조적이지 않은 데이터를 다루는데 사용되었지만, GNN은 그래프 데이터의 특징을 고려하여 노드 간의 관계를 학습할 수 있다. GNN은 이웃 노드들과의 상호작용을 통해 노드의 임베딩(Embedding)을 학습하고, 이를 기반으로 그래프 전체의 특징을 추출할 수 있다.

GNN의 기본 아이디어는 각 노드는 자신과 이웃 노드들의 정보를 종합하여 표현될 수 있다는 것이다. 이러한 종합된 정보는 그래프 내의 구조와 상호 작용을 고려한 특별한 형태의 임베딩으로 표현된다. 이 임베딩은 그래프 내에서 노드나 엣지의 분류, 예측, 군집, 연결 예측 등 다양한 그래프 기반 작업에 활용될 수 있다.

GNN은 최근 몇 년간 그래프 기계 학습 분야에서 큰 관심을 받고 있으며, 이를 통해 다양한 실제 응용 분야에서의 문제를 해결하는 데 활용되고 있다.
\subsection{GAT}

Graph Attention Network(GAT)는 그래프 데이터를 처리하기 위한 딥러닝 모델 중 하나로, Self-Attention 메커니즘을 사용하여 그래프 내의 노드 간 관계를 모델링하는 효과적인 방법이다. Self-Attention은 시퀀스 또는 시퀀스와 시퀀스 간의 관계를 모델링하는데 사용되는 메커니즘이다. 주로 Transformer 모델에서 처음 도입되었으며, 이후 다양한 자연어 처리 작업에서 큰 성공을 거뒀다. Self-Attention은 주어진 입력 시퀀스의 각 요소(단어 또는 토큰)에 대한 가중치를 계산하는 방법으로 작동한다. Self-Attention의 주요 단계는 다음과 같다:

1. 주어진 입력 시퀀스의 각 요소에 대해 쿼리(query), 키(key), 밸류(value)로 구성된 벡터를 생성한다. 이 때 벡터들은 학습 가능한 가중치로 인코딩된다. 

2. 쿼리 벡터는 다른 모든 키 벡터와의 유사도를 계산한다. 이 유사도는 점곱(내적, inner dot product)을 통해 얻어지며, 유사도를 소프트맥스 함수를 통과시켜 확률 가중치를 얻는다. 

3. 밸류 벡터에 확률 가중치를 곱해 가중 평균값을 계산하면, 입력 시퀀스 각 요소에 대한 새로운 표현을 얻을 수 있다. 

GAT는 Self-Attention 메커니즘을 그래프 데이터에 적용한 것이다. 그래프는 노드와 엣지로 구성된 네트워크이며, 각 노드가 다른 노드와의 관계를 가지고 있다. GAT는 이 관계를 고려하여 각 노드의 임베딩을 학습한다.
GAT의 주요 작동방식은 다음과 같다:

1. 각 노드는 자신과 이웃 노드의 정보를 가지고 있는 특성 벡터를 가진다. 이 특성 벡터는 초기화된 상태 또는 이전 레이어의 출력일 수 있다.

2. 각 노드에 대해 Self-Attention 메커니즘을 적용한다. 이 때 쿼리, 키, 밸류 벡터는 노드의 특성 벡터와 이웃 노드의 특성 벡터로부터 생성된다.

3. Self-Attention을 통해 각 노드는 이웃 노드와의 상대적인 중요성을 계산하고, 중요한 이웃 노드로부터 정보를 집계한다. 이 집계된 정보는 노드의 새로운 임베딩으로 사용된다.

4. Multi-Head Attention: GAT 모델은 여러 어텐션 헤드를 사용해 다양한 종류의 관계를 모델링한다. 각 어텐션 헤대는 서로 다른 가중치를 학습한다.

5. 잔차 연결: GAT 에서는 잔차 연결을 사용하여 훈련을 안정화시키고 기울기 소실 문제를 완화한다.

\begin{figure}[!ht]
\centering
   \includegraphics[width=5.5cm]{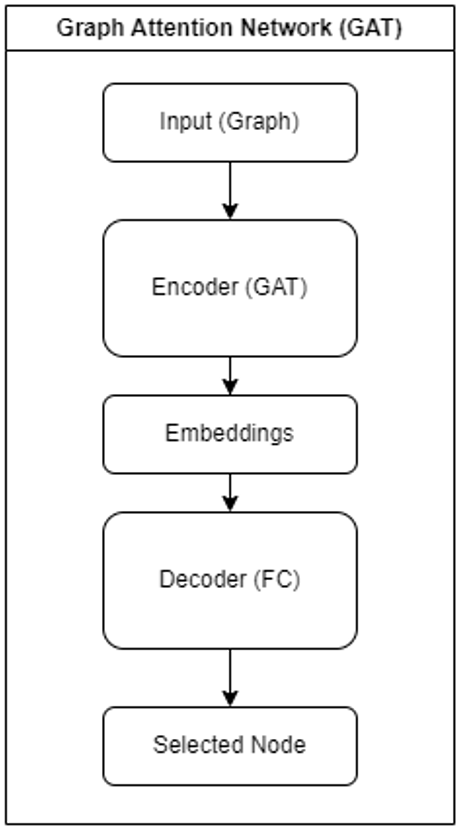}
\caption{GAT 흐름도}
\label{fig:008}
\end{figure}

결국, GAT를 사용하면 그래프 데이터 내에서 노드의 중요한 패턴과 관계를 감지하고 이를 활용하여 그래프 내에서 예측 및 분류를 수행할 수 있다. 이 모델은 다양한 분야에서 네트워크 데이터를 다루는데 유용하며, 자연어 처리, 컴퓨터 비전, 지식 그래프, 소셜 네트워크 분석 등 다양한 응용 분야에서 활발하게 연구되고 사용되고 있다. Figure \ref{fig:008}은 GAT 모델을 통해 이웃 노드 들의 가중치를 재정의한 뒤, 그 중 가장 높은 가중치를 가진 노드를 선택하는 경로탐색 GAT 의 예시를 보여준다.

\subsection{경로 탐색}

경로 탐색 알고리즘은 그래프나 네트워크와 같은 구조에서 출발점과 목적지 사이의 경로를 찾는 알고리즘이다. 목표는 출발점에서 목적지까지 가장 효율적인 경로를 찾거나, 특정 조건에 따른 경로를 찾는것이다. 경로는 그래프 노드와 엣지로 구성된 데이터 구조에서 정의되며, 엣지에는 가중치가 할당될 수 있다. 주요 경로 탐색 알고리즘으로는 깊이 우선 탐색(DFS), 너비 우선 탐색(BFS), 다익스트라 알고리즘, 벨만-포드 알고리즘, 크루스칼 알고리즘 등이 있다. DFS는 그래프를 깊이 우선으로 탐색하는 알고리즘으로, 목적지까지 가능한 한 깊숙히 들어가며 경로를 탐색한다. BFS의 경우 목적지로부터 멀리 있는 노드부터 차례대로 탐색한다. 다익스트라 알고리즘은 음의 가중치가 없는 그래프에 사용 가능하며, 우선순위 큐(Queue)를 사용하여 가장 짧은 거리를 가진 노드를 선택한다. 벨만-포드 알고리즘은 가중치가 있는 그래프에서 최단 경로를 찾는 알고리즘으로, 음의 가중치와 음의 사이클이 있는 그래프에서도 동작한다. 크루스칼 알고리즘은 최소 비용 신장 트리를 찾는 알고리즘이다. 모든 노드를 포함하면서 모든 엣지의 가중치 합이 최소가 되는 트리를 구성한다. 

경로 탐색 알고리즘은 그래프 구조와 문제의 성격에 따라 선택되며, 최적의 결과를 얻기 위해 적절한 알고리즘을 선택하는 것이 중요하다. 이러한 알고리즘은 네트워크 라우팅, 게임 인공 지능, 지리 정보 시스템, 소셜 네트워크 분석, 자율 주행 자동차 경로 계획 등 다양한 응용 분야에서 활용된다.

\subsection{강화학습}
강화학습(Reinforcement Learning)은 기계 학습의 한 분야로, 에이전트(Agent)가 환경(Environment)과 상호작용하며 보상(Reward)을 최대화하기 위한 행동(Action)을 학습하는 방법론이다. 이 방법은 에이전트가 주어진 환경에서 특정 상황에서 어떤 행동을 취애야 할지를 스스로 학습하게 하는데 중점을 둔다.

강화학습에서는 에이전트가 환경의 현재 상태를 관찰하고, 이 상태에 기반하여 선택 가능한 여러 행동 중에서 어떤 행동을 선택할지 결정한다. 에이전트가 선택한 행동에 따라 환경은 다음 상태로 전환되며, 이때 에이전트는 환경으로부터 보상을 받는다. 이러한 과정을 반복하면서 에이전트는 어떤 상태에서 어떤 행동을 취해야 더 큰 보상을 얻을 수 있는지를 학습하게 된다. 

강화학습의 기본 구성 요소는 다음과 같다.


에이전트는 의사결정을 하는 주체로, 학습 알고리즘이 적용되는 대상이다. 환경(Environment)는 에이전트가 상호작용하는 외부 시스템 또는 환경으로, 에이전트(Agent)의 행동(Action)에 의해 상태가 변하고 보상을 제공한다. 상태(State)는 현재 시점에서 환경의 상태를 나타내는 값으로, 에이전트의 의사 결정에 영향을 미친다. 행동(Action)은 에이전트가 환경에서 취할 수 있는 행동으로, 에이전트의 의사 결정 결과이다. 보상(Reward)는 에이전트가 환경으로부터 얻는 피드백으로, 특정 상태에서 취한 행동에 대한 평가를 나타낸다. 마지막으로 정책(Policy)는 상태에 따라 어떤 행동을 선택할지 결정하는 전략이나 알고리즘을 나타낸다. 

\begin{figure}[!ht]
\centering
   \includegraphics[width=5cm]{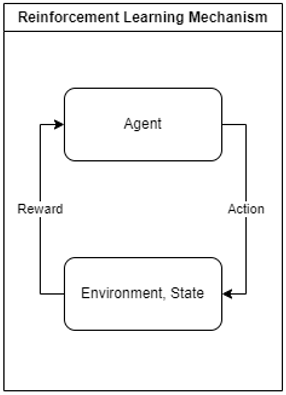}
\caption{Reinforcement Learning Mechanism}
\label{fig:007}
\end{figure}

강화학습은 에이전트가 보상을 최대화하기 위한 최적의 정책을 학습하는 것을 목표로 한다. 이를 위해 다양한 강화학습 알고리즘이 개발되어 상황에 따라 적절한 알고리즘을 선택하여 사용된다. 강화학습은 게임 플레이, 자율 주행 자동차, 로봇 제어, 자연어 처리 등 다양한 분야에서 응용되고 있다. Figure \ref{fig:007}은 강화학습의 기본적인 흐름을 보여준다. 에이전트의 행동이 상태를 변경하고, 변경된 상태가 보상으로 전환되어 에이전트에게 평가로 전해진다.

\subsection{Optimizer}
옵티마이저는 신경망 학습 과정에서 사용되는 최적화 알고리즘이다. 역전파를 통해 계산된 오차를 줄이기 위해 가중치를 업데이트할 때 옵티마이저가 사용된다. 일반적으로 경사하강법(Gradient Descent) 기반의 여러 옵티마이저가 사용된다. 이러한 옵티마이저들은 손실 함수(Loss Function)의 값을 줄이기 위해 모델의 가중치를 조정하는 역할을 한다. 몇 가지 대표적인 옵티마이저에는 확률적 경사하강법(SGD, Stochastic Gradient Descent), Adam, RMSprop 등이 있다. 

GAT(Graph Attention Network)은 그래프 데이터를 다루는 신경망 구조 중 하나로, 노드 간의 관계를 고려하여 그래프 데이터를 학습할 수 있다. Adam 옵티마이저가 GAT 환경에서 좋은 이유는 다음과 같다. 

GAT는 그래프의 노드 간에 가중치를 부여하는 방식을 사용한다. 이 때, 각 노드는 주변 노드와의 상호 작용에 따라 다른 가중치를 가질 수 있다. Adam 옵티마이저의 적응적 학습률은 이러한 가중치를 조정하고, 그래프의 복잡한 구조에 더 잘 적응할 수 있다. 또한 GAT는 그래프 내의 노드 간에 상이한 관계를 갖기 때문에, 각 노드의 학습 속도를 다르게 조정하는 것이 중요하다. Adam은 각 파라미터에 대한 독립적인 학습률을 제공함으로써 이를 가능하게 한다. 

GAT는 매우 복잡한 그래프 구조에서도 효과적으로 학습할 수 있는데, Adam 옵티마이저는 다양한 노드 간의 관계를 효과적으로 학습하고 최적화하기 위한 더 나은 수렴 속도를 제공할 수 있다. 이로 인해 GAT가 더 빠르게 수렴하고 더 나은 성능을 얻을 수 있다. Adam은 학습률 외에도 많은 하이퍼파라미터 튜닝 없이도 효과적으로 작동할 수 있는 경향이 있다. 이는 GAT와 같이 복잡한 모델 구조를 갖는 경우에 유용하다.
\section{관련 연구 (참고 논문)}
\label{sec:literature}
Kool등\cite{kool2018attention}은 라우팅문제를 최적화하는데 기계학습을 사용하는 아이디어를 제시한다. 전통적인 알고리즘에 비해 시간을 절약할 수 있는 방법이다. 어텐션 레이어를 기반으로한 모델을 제안하며, Pointer Network와 같은 기존 방법보다 이점을 가짐을 설명한다. 또한, 강화학습을 사용한 훈련 방법을 도입하여 greedy rollout에서 파생된 간단한 베이스라인을 사용하며, 이는 기존의 가치함수를 사용하는 것보다 효율적임을 입증한다.  

이 작업은 기계학습을 사용하여 여행하는 외판원 문제(TSP, Traveling Salesman Problem)에 대한 학습을 크게 개선하며, 노드 수가 최대 100개인 문제에 대해 거의 최적의 결과를 달성한다. 또한, 이를 확장해 차량 경로 문제(VRP), 오리엔티어링 문제(OP), 그리고 상품 수집 외판원 문제 (PCTSP)의 확률적 변형과 같은 문제를 해결하고자 한다. 그 결과, 기존의 알고리즘 성능과 유사한 값을 얻어낼 수 있었으며, 사용한 시간과 자원 측면에서 효율적임을 입증한다. 

이에 나아가, Drori등\cite{drori2020learning}은 그래프 문제에 대한 라우팅문제를 최적화하는데 집중한다. 해당 논문에서 제시하는 프레임워크는 상태, 행동 및 보상으로 정의된 단일 플레이어 게임으로 제시된 다양한 그래프 기반의 조합 최적화 문제를 해결한다. 이 프레임워크는 최소 신장 트리(MSP), 최단 경로, 외판원 문제, 차량 경로 문제를 포함하여 모든 그래프 문제를 처리한다. 라벨링이 되어 있지 않은 학습용 그래프 데이터 셋에서 강화학습을 사용하여 그래프 신경망을 훈련시키고, 훈련된 네트워크는 새로운 그래프 인스턴스에 대한 근사 솔루션을 선형 시간 내에 출력한다.

이 연구는 다양한 응용 분야에서 그래프 기반의 조합 최적화 문제를 해결하는데 딥러닝이 효율적임을 보여주며, GNN을 사용한 훈련과 강화학습을 결합하여 NP최적화 문제의 근사 솔루션을 찾을 수 있음을 보여준다. 특정 알고리즘에 집중하여 최적화 문제를 푸는 것이 아닌, 좀 더 일반적인 프레임워크로 작동하도록 학습하였으며, 기존 알고리즘 해결 방법과 비교해 충분히 효과적인 결과를 얻어낼 수 있었다.

\section{방법론}
\label{sec:methodology}
\subsection{Encoder}

실험에 사용한 모델은 크게 Encoder - Decoder - 강화학습 의 3단계로 나눌 수 있다. Encoder는 2층의 Attention Layer와 1층의 FeedForward Layer로 이뤄진다. Attention Layer는 노드의 특성을 업데이트하는 것이 주 목적이다. 초기 그래프 데이터가 Encoder로 들어오면, 각 그래프의 노드들이 가지는 특성값을 Attention Layer에서 임베딩한다. 이를 통해 노드의 값을 계산할 수 있다. 자기 자신의 가중치와 주변 노드의 가중치를 적절한 값으로 변환시킨 뒤, 새로운 노드 특성으로 업데이트한다. 활성함수로는 LeakyReLU를 쓰며, 정규화와 확률값으로 변경하기 위해 Softmax함수를 사용한다. 이후, 학습을 더 잘할 수 있도록 FeedForward Layer가 추가된다. FeedForward Layer를 추가하여 더 복잡한 함수와 관계를 학습할 수 있으며, Multi-head 에서 얻어온 Attention정보들을 변환하고, 조합하여 학습 성능을 높일 수 있다. Encoder는 결과적으로 업데이트된 노드 특성값 $v_i$ 를 Output으로 내보낸다.

\subsection{Decoder}

Encoder의 결과값이 Decoder의 입력값이 된다. 연결된 이웃 노드들 중 어느 노드를 선택할지 결정할 수 있도록 Decoder 단계에서는 업데이트된 특성 $v_i$ 를 통해 Attention Coefficients를 계산한다. 이는 주어진 노드 및 이웃 노드간의 상대적 중요성을 나타내는 값으로, 이 정보를 기반으로 다음에 선택할 노드를 결정한다. Attention Coefficients $a_{ij}$를 계산하는 식은 방정식 (\ref{eqn:AC})과 같다.

\begin{equation}\label{eqn:AC}
a_{ij}^{dec} = C tanh ((\Phi_1v_i)^T(\Phi_2v_j)/\sqrt{d_h})    
\end{equation}

Decoder는 각 노드에서 이웃 노드들의 Attention Coefficients 들을 출력한다. 이후 강화학습 단계 중 경로를 탐색할때, Decoder의 출력값을 확률로 활용하여 결정하게 된다.

\subsection{Reinforcement Learning}
\subsubsection{경로탐색}
먼저, 이 부분은 그래프의 각 노드를 하나씩 방문하며 최적의 경로를 찾는다. 이전에 방문한 노드를 기반으로, 현재 노드와 연결된 아직 방문하지 않은 노드를 찾는다. 만약 현재 노드에서 여러개의 미방문 노드가 있다면, 이를 나중에 다시 탐색할 수 있도록 분기점으로 기억한다.

그 다음, 하나의 미방문 노드를 선택하고, 딥러닝 모델(decoder)을 활용하여 다음 방문할 노드를 예측한다. 이때 선택된 노드와 해당 노드로 이동하는 확률을 계산하고, 가능한 경로의 확률을 기록한다. 이런 과정을 반복하여 모든 노드를 방문하면서 가능한 경로를 찾아낸다. 

이렇게 찾아낸 경로의 품질을 측정하기 위해 선택한 노드의 가중치를 누적하여 보상 값을 계산한다. 이 보상 값은 현재까지 선택한 경로의 길이나 비용을 나타내며, 이는 최적의 경로를 찾는 중요한 지표로 활용된다. 

마지막으로, 함수는 최정적으로 방문한 노드의 순서, 각 경로의 확률, 보상 값을 반환한다. 이 정보는 TSP 문제에서 최적의 경로를 찾는 데 도움이 되며, 이러한 정보를 토대로 모델의 학습과 경로 최적화가 이루어진다. 이렇게 함으로써 DFS 알고리즘과 딥러닝 모델이 함께 최적의 TSP 해답을 찾아가게 된다.

\begin{figure}[!ht]
\centering
   \includegraphics[width=8.5cm]{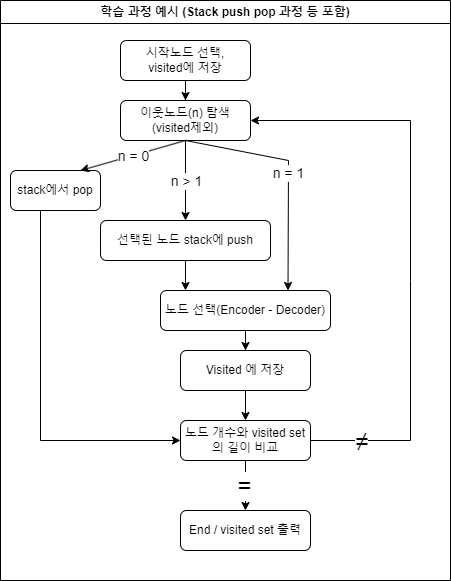}
\caption{경로탐색 Mechanism}
\label{fig:009}
\end{figure}

\begin{figure}
    \begin{minipage}{0.28\textwidth} 
        \centering
        \includegraphics[width=3cm]{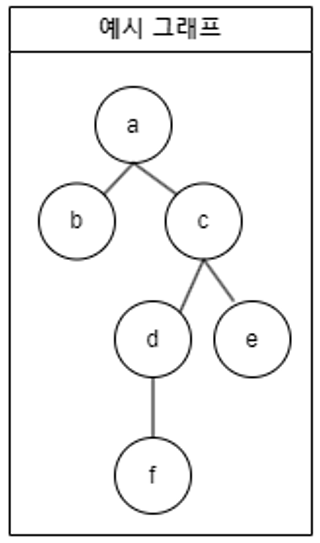}
        \caption{example graph}
        \label{fig:010}
    \end{minipage}
    \begin{minipage}{0.4\textwidth} 
        \begin{tabular}{lllll}
             선택노드& 이웃노드 & 다음노드 & visited\_node & stack \\ \hline\hline
             a & b,c & b & \{a,b\} & \{a\} \\ \hline
             a & c & c & \{a,b,c\} & \{\} \\ \hline
             c & d,e & d & \{a,b,c,d\} & \{c\} \\ \hline
             d & f & f & \{a,b,c,d,f\} & \{c\} \\ \hline
             c & e & e & \{a,b,c,d,f,e\} & \{\} \\ \hline
        \end{tabular}
        \caption{경로 탐색 과정 예시}
        \label{tab:001}
        \end{minipage}
\end{figure}

Figure \ref{fig:009}는 이 연구에서 제안된 경로탐색 알고리즘을 보여주고 있다. 그리고 이의 특정한 예시인 Figure \ref{fig:010}의 상황에서 경로 탐색 과정 예시를 표시해보면 table \ref{tab:001}와 같이 표현될 수 있다. 그 설명은 아래와 같다.

먼저, 초기 노드가 ``a"로 선택된다. 그 다음에는 초기 노드의 이웃 노드 중에서 확률이 높은 노드가 선택된다. 이 경우, 이웃 노드 ``b"가 선택된다. 그런 다음, 선택된 노드 ``b"가 ``visited node" 리스트에 추가되고, 초기 노드 ``a"는 스택에 쌓인다. 이때, 노드 ``a"의 이웃 노드가 2개 이상인 경우 스택에 쌓이게 된다.

다음으로, 노드 ``b"의 이웃 노드 수가 0개인 것을 확인하게 되면, 스택에 있는 노드 ``a"가 pop되어 꺼내지게 된다. 그러면 노드 ``a"에서 이웃 노드를 다시 찾게 되고, 이러한 과정을 반복하면서 경로가 탐색된다. 이 과정을 표로 나타낸 것이 아래의 표이다.

\subsubsection{학습 방법}
주어진 그래프 데이터에 대해 GAT 모델을 사용하여 노드의 임베딩을 생성하고, 이를 기반으로 TSP 경로를 찾기 위한 디코딩을 수행한다. 훈련 루프는 여러 그래프에 대해 반복되며, 각 epoch에 대해서는 다음과 같은 단계를 거친다.

먼저, 각 그래프에서 무작위로 시작 노드를 선택하고, 해당 노드의 임베딩을 GAT 모델을 사용하여 계산한다. 선택한 시작 노드를 기반으로 `decode\_all' 함수를 사용하여 TSP 경로를 확장하고, 이에 대한 보상을 계산한다. 또한, 베이스라인 모델을 사용하여 경로의 보상 기준값을 계산한다. 

계산된 보상과 베이스라인 보상 간의 차이를 최소화하기 위한 손실 함수를 계산한다. 손실을 최소화하며 GAT 모델의 파라미터를 업데이트한다. 주기적으로 베이스라인 모델을 업데이트하여 학습의 안정성을 유지한다. n개의 그래프에서 계산된 평균 손실과 보상 값을 기록하고, 훈련이 끝난 후에는 전체 훈련 기간을 측정하여 기록한다.

이러한 과정을 통해 GAT 모델은 TSP 문제에 대한 최적의 경로를 학습하며, 훈련을 통해 손실을 최소화하여 모델의 성능을 향상시킨다.

\section{실험 및 결과}
\label{sec:experiment}
\subsection{무작위 가중치 그래프 생성}
우리의 실험에서는 무방향 그래프를 생성하는 방법으로 랜덤 가중치 그래프를 활용했다. 이를 위해 Python 프로그래밍 언어를 사용하여 NetworkX 라이브러리와 PyTorch를 결합한 함수를 개발했다. 우선, 주어진 노드의 수(`num\_nodes')와 에지의 수(`num\_edges')에 따라 빈 무방향 그래프를 생성하고, 각 노드에 0과 1 사이의 랜덤한 가중치를 부여했다. 이러한 가중치는 노드의 속성으로 저장되고, 동시에 노드의 특징 벡터로 표현되었다. 다음으로, 임의의 시작 노드를 선택하고 해당 노드와 나머지 노드 사이에 에지를 형성하여 그래프를 연결했다. 추가적인 랜덤 에지가 필요한 경우, 남은 에지의 수를 충족시키기 위해 랜덤한 출발 노드와 도착 노드 간에 에지를 추가했다. 최종적으로는 그래프의 인접 행렬을 계산하고, 이를 3차원 텐서로 변환하여 모델 학습에 사용할 수 있는 형태로 만들었다. 또한, 그래프의 랜덤한 출발 노드를 선택한 인덱스 `j'도 함께 반환된다. 이러한 방법을 통해 무작위 가중치를 가진 무방향 그래프를 효과적으로 생성하고 활용할 수 있었다.  

\begin{figure}[!ht]
\centering
   \includegraphics[width=8.5cm]{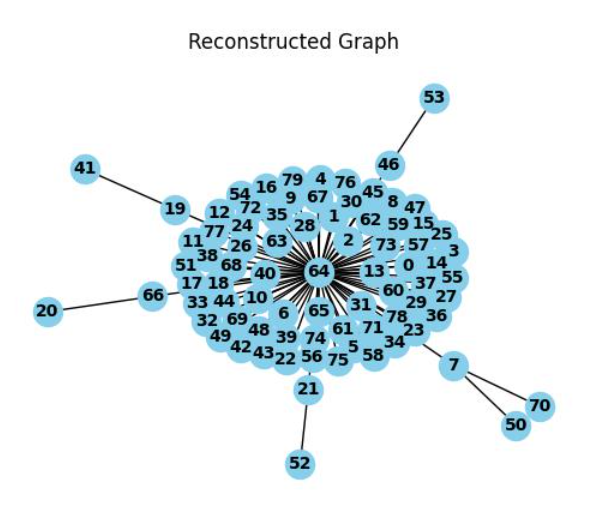}
\caption{Generated Random Graph.}
\label{fig:003}
\end{figure}

우리의 연구에서 실제 네트워크의 상태를 더 정확하게 반영하기 위해 무방향 그래프를 생성할 때 트리와 유사한 구조를 고려했다. 실제로 많은 실제 네트워크는 트리 구조에 가까운 경향이 있으며, 이러한 구조는 복잡한 네트워크 상황을 단순화하면서도 중요한 특성을 보존할 수 있다. 따라서, 우리가 생성한 무방향 그래프에서 트리와 유사한 특성을 얻기 위해 그래프의 엣지의 수를 노드의 수보다 조금만 더 많게 설정을 했다. 이렇게 함으로써 각 노드는 대부분의 경우 다수의 이웃 노드와 연결되어 있으며, 네트워크가 보다 현실적인 상황을 모방하도록 하였다. 이러한 조정된 그래프 구조는 우리의 연구에서 모델의 성능을 높이는 데 기여하였으며, 실험 결과에서도 이를 확인할 수 있었다. 아래의 Figure \ref{fig:003} 은 노드 100개의 랜덤 그래프를 생성한 결과이다.

\subsection{실험 결과}

\begin{figure}[!ht]
\begin{subfigure}{0.5\textwidth}
    \centering
    \includegraphics[width=7cm]{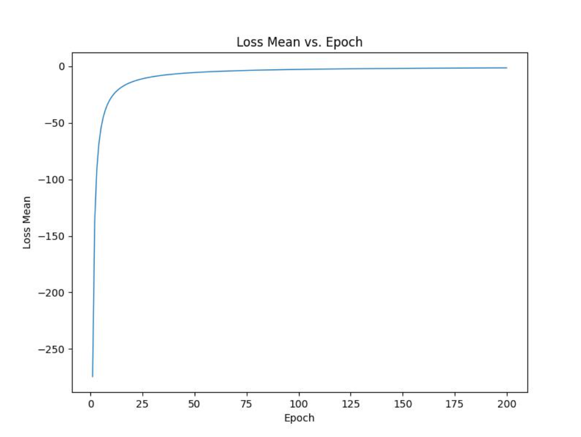}
    \caption{Model Loss Graph.}
    \label{fig:001}
\end{subfigure}%
\begin{subfigure}{0.5\textwidth}
    \centering
    \includegraphics[width=7cm]{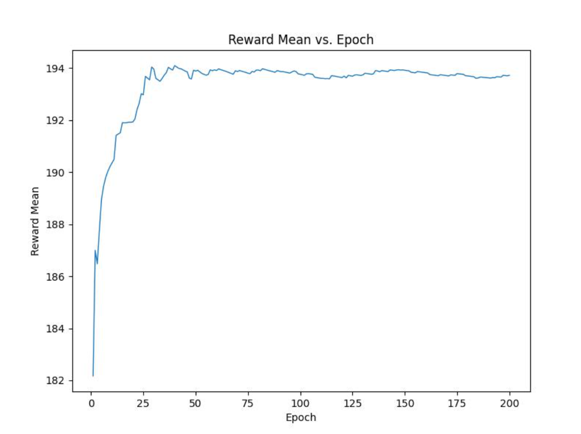}
    \caption{Model Reward Graph.}
    \label{fig:002}
\end{subfigure}
\caption{Model Loss and Reward Graphs}
\end{figure}

본 실험에서는 제안한 GAT 기반 경로 추론 모델의 성능을 평가했다. 실험에서는 먼저 모델이 학습 데이터에 대해 어떻게 수렴하는지를 관찰하기 위해 Loss 값의 변화를 분석하였다. Figure \ref{fig:001} 은 학습 과정에서의 Loss 값의 추이를 나타내고 있다. 모델은 초기 에폭에서부터 Loss 값을 감소시키며 약 30에서 50 번의 에폭에서 안정적으로 수렴하는 것을 확인할 수 있었다. 

또한, Figure \ref{fig:002} 는 GAT 기반 모델을 사용하여 무작위로 생성된 무방향 그래프에 대한 경로 추론 결과를 나타내고 있다. 각 그래프는 각 끝 노드에 대한 Reward 값은 GAT 기반 모델의 성능을 나타낸다. 실험 결과에서 GAT 기반 모델이 공격 경로를 효과적으로 추론할 수 있음을 확인 할 수 있었다. 

\subsection{경로 추론 검증 방법}
GAT 기반 경로 추론 모델의 정확성을 검증하기 위해 Brute Force 방법을 활용하였다. Brute Force를 통해 시작 노드에서 i번 노드까지 갈 수 있는 Attack Path Score를 구하는 과정은 다음과 같은 pseudo code로 표현된다.

\include{pythonlisting}

\begin{algorithm}[H]
\SetAlgoLined

  \If{next\_node == end\_node}{
  Calculate Attack Path Score\;
  \If{current\_path Attack Path Score is greater than Saved Attack Path Score}{
   Update Saved Attack Path Score\;
   }{
   \For {neighbors of next\_node} {\If{neighbor not in current\_path}{old path = current\_path\;DFS()}} 
    }}

\caption{DFS(next\_node, end\_node, current\_path)}
\end{algorithm}

\begin{algorithm}[H]
\SetAlgoLined
\KwResult{Highest Attack Path Score for each End Node}

 \For {end\_node in nodes} {DFS()}
\caption{Brute Force}
\end{algorithm}
\begin{figure}[!ht]
\centering
   \includegraphics[width=8.5cm]{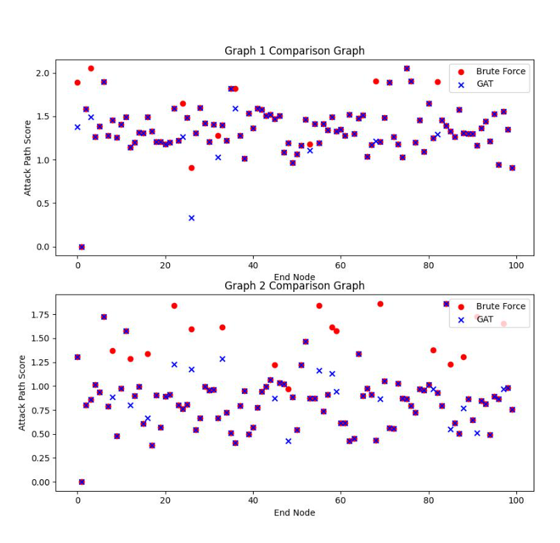}
\caption{Graph Comparison.}
\label{fig:004}
\end{figure}
위 코드를 기반으로 우리는 오리지널 그래프와 GAT로 재구성한 그래프의 Attack Path Score를 비교하였다. 이를 통해 우리는 재구성된 그래프가 오리지널 그래프와 얼마나 유사한지를 확인하였고, 경로를 잘 찾았는지를 확인하였다. 아래 Figure \ref{fig:004}는 이를 시각화한 것이다.

이를 통해 GAT 기반 경로 추론 모델이 오리지널 그래프의 특성을 효과적으로 학습하고, 정확한 Attack Path Score를 추론할 수 있음을 확인하였다.

\section{결론}
\label{sec:conclusion}
본 기술문서에서 인공지능 기반의 공격 그래프 생성에 관한 내용을 기술하였다.
공격 그래프 생성을 위한 인공지능 모델은 encoder와 decoder의 구조를 지녔다.
Encoder는 그래프 데이터를 입력으로 받아 그래프의 각 노드의 가중치(임베딩 값)를 자체적으로 계산해 출력했으며, decoder는 encoder가 출력한 노드의 가중치를 바탕으로 노드와 엣지를 선택했다.
선택한 노드와 엣지로 구성된 부분 그래프에 대해 경로 탐색을 수행하며 강화학습 기법으로 인공지능 모델을 학습하였다.
인공지능 모델을 학습하기 위해 무작위 가중치 그래프를 생성해 학습 데이터 및 테스트 데이터로 사용하였으며, 모델이 잘 학습되었는지 확인하기 위해 loss 값과 reward 값을 확인하였다.
Loss와 reward 값이 수렴하는 것으로 미루어 인공지능 모델이 잘 학습됨을 확인했다.
학습된 인공지능 모델을 좀더 검증하기 위해 brute force 방법으로 모든 노드 간의 attack path를 계산하여 그 결과를 인공지능 모델의 결과와 비교하였다.
두 결과는 비슷했으며 동일한 경우도 있었다.
이를 통해 본 기술문서의 인공지능 모델이 생성한 공격 그래프가 기존 방법으로 생성한 공격 그래프와 비슷한 것을 확인할 수 있었다.

\bibliographystyle{plain}
\bibliography{main}

\begin{figure}[b] 
\centering
\subfloat{
\includegraphics[width=0.35\linewidth]{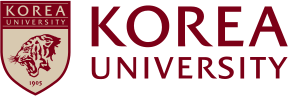}
}
\centering
\subfloat{
\includegraphics[width=0.35\linewidth]{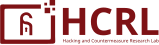}
}

\end{figure}

\end{document}